# Magnetocapacitance and exponential magnetoresistance in manganite-titanate heterojunctions

N. Nakagawa, M. Asai, Y. Mukunoki, T. Susaki, and H. Y. Hwang[a]

Department of Advanced Materials Science, University of Tokyo, Chiba 277-8561, Japan

ABSTRACT

We present a rectifying manganite-titanate heterojunction exhibiting a magnetic field tunable depletion layer. This creates a large positive magnetocapacitance, a direct measure of the field-induced reduction of the effective depletion width across the junction. Furthermore, the reduction of the junction barrier shifts the forward bias characteristics, giving exponentially-enhanced differential magnetoresistance, occurring despite the absence of a spin filter. These results provide a unique probe of a Mott insulator/band insulator interface, and further suggest new electronic devices incorporating the magnetic field sensitivity of these strongly correlated electron materials.

There is a continual search for new methods and materials to utilize magnetic response in electronic devices, ranging from dilute magnetic semiconductors,[1] metal superlattices,[2] magnetic oxides,[3] as well as hybrid structures among them. Perovskite manganites exhibit strong electron-spin coupling, manifesting dramatic magnetoresistance near the simultaneous magnetic and metal-insulator transition,[3] as well as a fully spin-polarized ground state.[4] The later feature has been explored in spin-polarized transport at grain boundaries[5] and thin film tunnel junctions.[6] Electron-doped $SrTiO_3$ has long been used as a semiconducting element in rectifying junctions.[7,8] Oxygen vacancies, as well as substitutional doping such as Nb on the Ti-site in $SrTiO_3$, generate conduction electrons.[9] Recently, diode characteristics have been demonstrated in manganite-titanate junctions,[10-13] where the manganite hole concentration was initially minimized to resemble semiconductor *p-i-n* or *p-n* junctions, corresponding to bulk insulating concentrations.

Ferromagnetic metallic manganites are formed by doping holes in the antiferromagnetic Mott insulator $LaMnO_3$, with the metal-insulator transition occurring for ~ 17 % hole doping in $La_{1-x}Sr_xMnO_3$.[3] We have found that rectification can be observed even for much higher hole concentrations, and focus here on junctions using $La_{0.7}Sr_{0.3}MnO_{3-\delta}$ to vary across the ferromagnetic state. For $La_{0.7}Sr_{0.3}MnO_{3-\delta}$/Nb:$SrTiO_3$ junctions, we find a significant increase in the junction capacitance in an applied magnetic field. Corresponding to this decrease in the depletion width, the reduction of the current barrier gives rise to exponential differential magnetoresistance. The modification of the electronic structure at the heterointerface by magnetic field is enabled by the unusual features arising from strong charge-spin coupling, giving a direct probe of the correlated electron equivalent of semiconductor band bending.

We grew $La_{0.7}Sr_{0.3}MnO_{3-\delta}$ films (80 nm thick) on Nb 0.01 weight % doped $SrTiO_3$ (001) single crystal substrates using pulsed laser deposition. A KrF excimer laser was used with a repetition rate of 4 Hz with laser fluence at the target surface of ~ 3 J/cm². The films were grown at 700-750 °C in an oxygen partial pressure of either 250 mtorr (stoichiometric films) or $10^{-3}$ torr (oxygen deficient films). $La_{0.7}Sr_{0.3}MnO_3$ is well lattice-matched to $SrTiO_3$ (< 1 % pseudocubic mismatch), and the single crystal films exhibited typical magnetic and transport properties for both stoichiometric[3] and oxygen deficient films,[14] as exemplified by the resistivity shown in Fig. 1(a). Ohmic contacts were prepared by evaporating Au and Al on $La_{0.7}Sr_{0.3}MnO_3$ and $SrTiO_3$, respectively, and the junctions were scribed to dimensions ranging from 0.2-2 mm². The junctions were quite robust, with 100% initial yield and good scaling to area.

Figures 1(c) and (d) shows the temperature dependent current-voltage characteristics of the oxygen deficient and stoichiometric manganite-titanate junctions, with the polarity given in Fig. 1(b). The forward bias region can be described by a shift to higher voltage with decreasing temperature, while the reverse bias region exhibits a breakdown feature which appears at lower reverse bias with decreasing temperature. Below ~200 K, the usual temperature dependence for semiconductor $p$-$n$ or Schottky diodes is not observed for the current density $J = J_S(T)(\exp[qV/kT] - 1)$, where $J_S$ is the saturation current density, $q$ is the electronic charge, and $k$ is the Boltzmann constant. Both the film and doped substrate are metallic to low temperature, with no evidence for carrier freezeout, rendering the standard semiconductor picture not directly applicable, even neglecting the effects of electron correlations.[15]

Figure 2(a) shows the magnetic field dependence for the $La_{0.7}Sr_{0.3}MnO_{3-\delta}$ junction at 10 K, with the field applied perpendicular to the planar junction. The magnetic field induces a large shift of the forward bias characteristics to lower voltage, with negligible effect on the reverse bias region after the initial application of the magnetic field (see Fig. 3). It is reasonable to suspect the manganite as the origin of the magnetoresistance, given the relative insensitivity to magnetic field for conventional nonmagnetic semiconductor diodes. Indeed, we have confirmed that gold-SrTiO$_3$ Schottky junctions display negligible magnetoresistance on this scale at all temperatures (data not shown). Figure 2(b) shows the much smaller magnetic field dependence of the stoichiometric manganite junction at 10 K.

In order to probe the effect of a magnetic field, we have measured the magnetocapacitance of the La$_{0.7}$Sr$_{0.3}$MnO$_{3-\delta}$-titanate junction. Figure 3 shows a central result of this study: *the magnetic field reduces the effective depletion width, exponentially enhancing the junction magnetoresistance*. Figure 3(a) shows the low-frequency junction capacitance at zero bias at 10 K as a function of applied magnetic field. Similar with many magnetic devices, the initial effect is enhanced, in which the device was cooled from high temperature in zero applied field. After exposure to high magnetic field, the hysteretic capacitance is very reproducible for subsequent measurements. Figure 3(b) shows the differential conductance $G = dJ/dV$ for the same device at 10 K at a bias of 0.76 V (indicated by arrows in Fig. 2(a)), which, plotted on a logarithmic scale, directly mirrors the capacitance.

Figures 3(a) and (b) demonstrate that the magnetoresistance derives from reduction of the junction barrier, without spin selection of the electrons that traverse the interface. Due to the exponential voltage dependence, $G$ increases by almost two orders of magnitude in 8 T, while the capacitance increases 33 %. With increasing temperature, both the junction magnetocapacitance and magnetoresistance diminish, maintaining this direct relationship. The stoichiometric manganite-titanate junction, by contrast, has little magnetocapacitance at all temperatures (data not shown). For both La$_{0.7}$Sr$_{0.3}$MnO$_{3-\delta}$ and La$_{0.7}$Sr$_{0.3}$MnO$_3$, the positive magnetocapacitance is opposite that expected from Zeeman splitting of the manganite *d* bands.[16]

In that case, negative magnetocapacitance arises due to an increase in the net screening length via spin-splitting.

In Figs. 4(a) and (b), we examine the effect of a magnetic field on the reverse-bias voltage dependence of the junction capacitance at 300 K and 10 K. For a *p-n* heterojunction, the junction capacitance $C$ should have the form

$$\frac{1}{C^2} = \frac{2}{q}\left[\frac{1}{\varepsilon_{STO} N_{STO}} + \frac{1}{\varepsilon_{LSMO} N_{LSMO}}\right][V_{bi} - V] \qquad (1)$$

where $\varepsilon_{STO}$ and $\varepsilon_{LSMO}$ are the permittivity of $SrTiO_3$ and $La_{0.7}Sr_{0.3}MnO_{3-\delta}$, $N_{STO}$ and $N_{LSMO}$ are the two carrier densities near the interface, and $V_{bi}$ is the built in potential across the heterojunction.[15] At 300 K, the data are consistent with the usual semiconductor form, resulting in a straight line for $1/C^2$ versus bias voltage. At 10 K, however, the complexity of this heterointerface becomes apparent. The nonlinearity arises at least in part due to the large electric field dependence of $\varepsilon_{STO}$, which has been well established.[17] The systematic magnetocapacitance shift in Fig. 4(b) demonstrates the correspondence between magnetocapacitance and differential magnetoconductance observed in Fig. 3 is robust with respect to applied bias.

Isolating a single origin for the magnetic field dependent depletion layer is nontrivial. Equation 1 and the junction current equation enforce continuity of the electric displacement at the heterointerface, coupling terms on either side of the junction. Futhermore, many of the unique features of manganites arise precisely due to strong coupling of the conduction electrons, local moments, and the lattice, resulting in near degeneracy of competing ground states.[3] This competition results in dramatic changes by varying temperature or magnetic and electric fields. Despite these complications, a simplified picture can be deduced. The subdued temperature dependence of the junction transport is inconsistent with diffusive or thermionic processes, suggesting a tunneling mechanism at low temperatures. For tunneling, the background dielectric constants do not appear in the WKB approximation, but enter via the image potential.[18] Assuming the magnetic field does not alter this term, the magnetic field simply reduces the barrier area, giving the linear magnetocapacitance and exponential magnetoresistance we observe.

We note that a simple magnetic field induced delocalization transition cannot explain our results, but rather a significant rearrangement of the density of states is required. Two effects are relevant – 1) Except for the case of a ¼ filled band, changes in spin order by temperature or magnetic field lead to a sizable shift in the chemical potential by changing the electronic bandwidth;[19] 2) The magnetic field increases the net density of states, a remnant of the metal-insulator transition which can be induced by field at low doping, creating a coherent band at the chemical potential.[3] Together these effects give a reduction of the depletion width and transport barrier under an applied magnetic field. Although our results bear similarities to a recently proposed magnetic/nonmagnetic p-n junction, in that case the origin of the magnetoresistance is the Zeeman splitting of the nonmagnetic minority carrier band,[20] not the change in depletion width we have observed.

These results have bearing on magnetotransport in manganite-titanate-manganite tunnel junctions, which have been extensively studied.[3] The manganite-titanate junctions studied here can be considered half of a tunnel junction, even in the detailed band alignment, as the chemical potential of insulating $SrTiO_3$ is pinned near the conduction band by residual impurities or vacancies. Studies of the magnetoresistance in tunnel junctions have focused on the spin-dependent density of states for magnetically aligned and anti-aligned electrodes, assuming a constant barrier potential.[21] Here we find that the barrier itself can be strongly modulated by the magnetic field. The application of these results to grain boundary interfaces is less direct, although surface depletion has been proposed as a model of the grain boundary junction.[22]

Current efforts to enhance the performance of manganite tunnel junctions have focused on improving the perfection of the interface.[23] This study suggests a different approach. A marginally unstable, thin magnetic layer at a heterojunction may provide an indirect but effective coupling between an applied magnetic field and junction transport.

We acknowledge partial support from NEDO's International Joint Research Program. N.N. acknowledges partial support from QPEC, Graduate School of Engineering, University of Tokyo.

**Footnote:**

a) Also at: Japan Science and Technology Agency, Kawaguchi, 332-0012, Japan; electronic mail: hyhwang@k.u-tokyo.ac.jp

**Figure Captions:**

**FIG. 1.** (a) Temperature dependent resistivity for $La_{0.7}Sr_{0.3}MnO_{3-\delta}$ and $La_{0.7}Sr_{0.3}MnO_3$ films in 0, 4, and 8 tesla applied field. (b) Schematic illustration of the junction device and polarity. (c) Temperature dependence of the current voltage characteristics of the $La_{0.7}Sr_{0.3}MnO_{3-\delta}$ junction. (d) Temperature dependence of the current voltage characteristics of the $La_{0.7}Sr_{0.3}MnO_3$ junction.

**FIG. 2.** (a) Magnetic field dependence of the $La_{0.7}Sr_{0.3}MnO_{3-\delta}$ junction forward bias characteristics at 10 K. (b) Magnetic field dependence of the $La_{0.7}Sr_{0.3}MnO_3$ junction forward bias characteristics at 10 K.

**FIG. 3.** Magnetic field dependence of the $La_{0.7}Sr_{0.3}MnO_{3-\delta}$ junction characteristics at various temperatures. (a) Zero-bias junction capacitance for magnetic field sweeps across ± 8 tesla. (b) Differential conductance for magnetic field sweeps across ± 8 tesla. The junction bias was fixed to correspond to 0.02 A/cm$^2$ at 8 tesla (indicated by arrows for 10 K in Fig. 2(a)), corresponding to 0.76 V, 0.71 V, and 0.59 V at 10, 40 and 100 K, respectively. The initial field sweep 0-8 tesla at 10 K was taken after cooling in zero applied magnetic field for both panels (a) and (b).

**FIG. 4.** (a) 1/C$^2$ as a function of bias voltage at 300 K in 0 and 8 tesla field. (b) 1/C$^2$ as a function of bias voltage at 10 K in varying magnetic field. Note the strong curvature arising from the large electric field dependence of $\varepsilon_{STO}$.

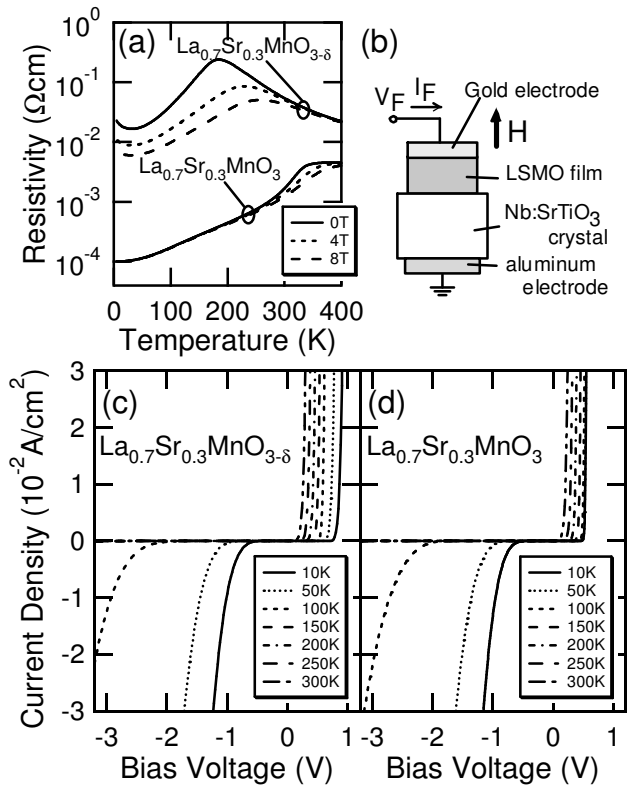

Fig 1. N. Nakagawa *et al.*

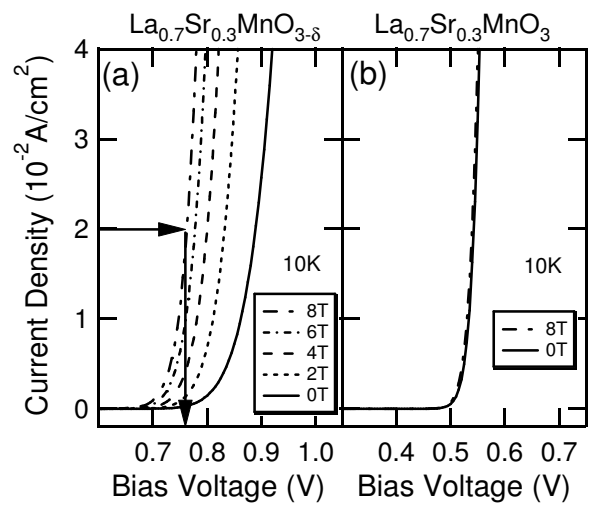

Fig 2. N. Nakagawa *et al.*

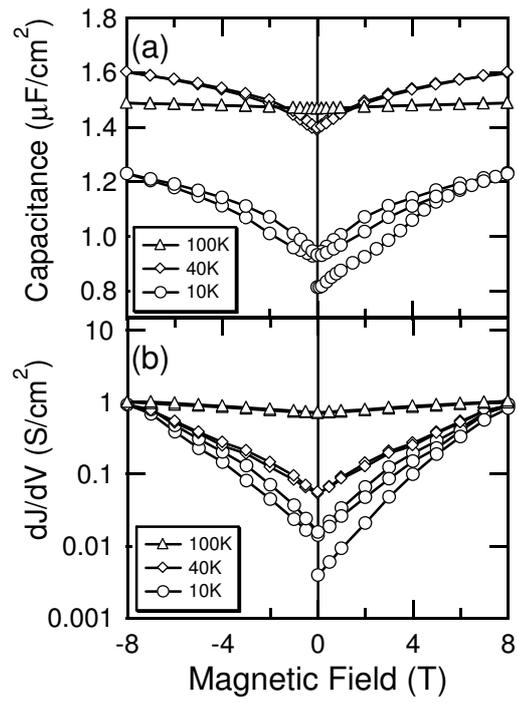

Fig 3. N. Nakagawa *et al.*

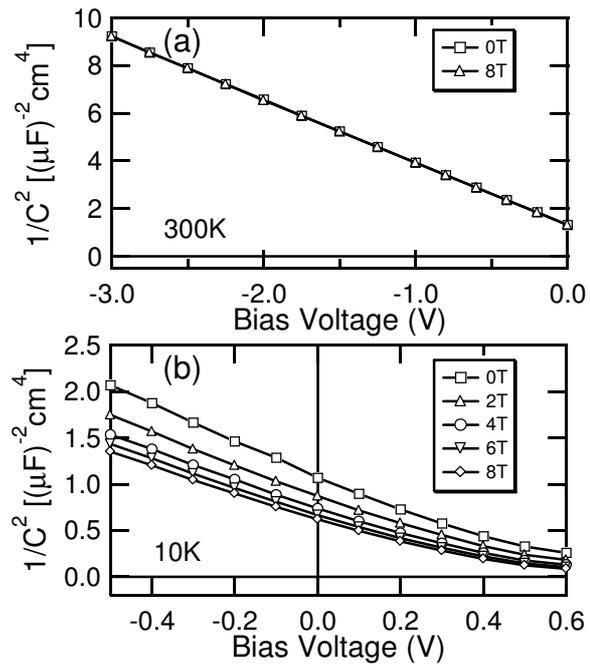

Fig 4. N. Nakagawa *et al.*